\documentclass[12pt]{iopart}
\usepackage{iopams}  
\usepackage{setstack}
\usepackage{braket}

\usepackage{textcomp}
\usepackage{graphicx}
\usepackage{color}
\usepackage[colorlinks=true,citecolor=blue,urlcolor=blue,linkcolor=blue]{hyperref}
\usepackage[all]{hypcap}

\newcommand{\vect}[1]{\boldsymbol{#1}}

\begin{document}

\title[Doppler narrowing, Zeeman and laser beam-shape effects in $\Lambda$-EIT of Rb$^{85}$]{Doppler narrowing, Zeeman and laser beam-shape effects in  $\Lambda$-type Electromagnetically Induced Transparency on the $^{85}$Rb D2 line in a vapor cell}

\author{L.~Ma$^1$,G.~Raithel$^{1}$}

\address{$^1$Department of Physics, University of Michigan, Ann Arbor, Michigan 48109}

\ead{lukema@umich.edu}

\date{\today}

\begin{abstract}
We study $\Lambda$-type Electromagnetically Induced Transparency (EIT) on the Rb D2 transition in a buffer-gas-free thermal vapor cell without anti-relaxation coating. Experimental data show distinguished features of velocity-selective optical pumping and one EIT resonance. The Zeeman splitting of the EIT line in magnetic fields up to 12 Gauss is investigated. One Zeeman component is free of the first-order shift and its second-order shift agrees well with theory. The full width at half maximum (FWHM) of this magnetic-field-insensitive EIT resonance is reduced due to Doppler narrowing, scales linearly in Rabi frequency over the range studied, and reaches about 100~kHz at the lowest powers. These observations agree with an analytic model for a Doppler-broadened medium developed in Ref.~\cite{PhysRevA.66.013805,7653385}. Numerical simulation using the Lindblad equation reveals that the transverse laser intensity distribution and two $\Lambda$-EIT systems must be included to fully account for the measured line width and line shape of the signals. Ground-state decoherence, caused by effects that include residual optical frequency fluctuations, atom-wall and trace-gas collisions, is discussed.
\end{abstract}




\section{Introduction}

Light-matter quantum-state entanglement and manipulation have been a research focus in the condensed-matter and AMO communities for many years~\cite{PhysRevLett.117.077403,PhysRevLett.86.783}.  In contrast to a laser cooled atomic sample, the atoms in a gaseous vapor phase pose a wide range of Doppler-shifts, which leads to the famous hole burning and Lamb dip phenomena. Recently, successful implementation of EIT spectroscopy~\cite{PhysRevLett.74.666,PhysRevLett.98.113003} in gaseous samples have sparked new ideas for making quantum enabled devices such as atomic-optical clocks~\cite{Knappe:01}, sensitive motion sensors~\cite{PhysRevLett.124.093202}, magnetometers~\cite{PhysRevA.83.015801,PhysRevA.82.033807} ,  microwave sensing devices~\cite{6910267}, single photon optical switches~\cite{Dawes672}, quantum memories~\cite{PhysRevLett.105.153605,vanderWal196} etc. All of those advancements rely on high-level coherent control of the interaction between a gaseous medium and light.

The laser induced atomic coherence can be fragile, as it is affected by various decoherence processes, including optical pumping, collision or diffusion, and power broadening. For applications such as atomic frequency standards and precision magnetometers that utilize vapor cells, atomic coherences arising from Coherent Population Trapping (CPT) have been reported in detail in terms of line width and line shape~\cite{PhysRevA.69.024501,6765452,Figueroa:06}. In these systems, collisions between the probed atoms and buffer gas atoms in the vapor cell contribute significantly to the homogenous line width. Depending on the buffer gas pressure and collision conditions, this broadening can be larger than the Doppler width~\cite{PhysRevA.58.2345,PhysRevA.63.043813}, yet the CPT resonances remain very narrow. Under such circumstances, the CPT line shape and width are determined by atom diffusion and local light intensity~\cite{PhysRevLett.96.043601,8698263,PhysRevA.69.024501}.

EIT experiments in thermal vapor cells without buffer gas~\cite{PhysRevLett.124.093202,PhysRevA.101.013821,PhysRevLett.116.013601,PhysRevLett.105.153605,Wang2018,OC8434,PhysRevA.65.023806,PhysRevA.79.013808} also have received considerable attention, as Ref.~\cite{PhysRevA.65.023806} points out that the EIT resonance is a unique product of the light-atom coherences that can be experimentally measured and can provide us an opportunity to better understand the influence of different decoherence processes.

\begin{figure}\label{fig1}
\includegraphics[width=0.7\textwidth]{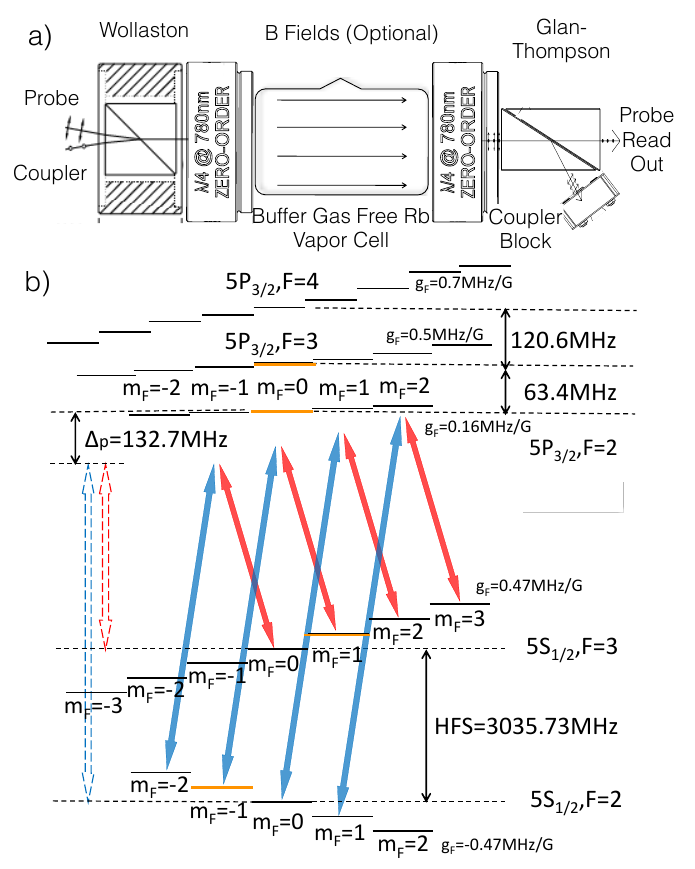}
\caption{(a) Illustration of the experimental setup featuring co-propagating coupler and probe beams in $\sigma^+$-$\sigma^-$ polarization configuration passing through a Rb vapor cell of 25~mm in diameter and 70~mm in length. A variable uniform magnetic field can by applied via a solenoid with inner diameter 33~mm, outer diameter 44~mm, and length 190~mm (not shown). (b) Zeeman level diagram of relevant states and transitions in the given polarization configuration in a magnetic field for the EIT study. The blue and red arrows correspond to transitions driven by the coupler and probe lasers. The scheme breaks up into four $\Lambda$ systems that correspond to individual, Zeeman-shifted EIT lines. The $\Lambda$ system with zero first order Zeeman shift has been highlighted with orange energy-level bars.}
\end{figure}

Expanding on previous work in~\cite{OC8434,PhysRevA.65.023806,PhysRevA.79.013808}, we present EIT measurements for a $\Lambda$-type system on the $^\mathrm{85}\mathrm{Rb}$ D2 transition in a buffer-gas-free vapor cell without anti-relaxation coating. An illustration of the experimental setup can be found in Figure~\ref{fig1}(a). We first exhibit the reduced (saturated) and enhanced absorption lines caused by velocity-selective optical pumping on the ground- and excited-state hyperfine structure, as well as the location of the EIT resonance within the overall spectrum. For our study of the EIT line width, we select a resonance with zero first-order Zeeman shift and eliminate inhomogeneous line broadening and pulling effects by lifting the Zeeman degeneracy with an in-situ calibrated spatially homogeneous magnetic field. We demonstrate that the width of the magnetic-field-insensitive EIT line varies linearly as a function of the coupling-laser Rabi frequency. Our results confirm the theoretical prediction outlined in Ref.~\cite{PhysRevA.66.013805,7653385} for a Doppler broadened sample. Further, a numerical simulation in which we include the laser intensity profile shows improved fitting to our data. This aspect is not fully accounted for in previous theoretical~\cite{PhysRevA.66.013805,7653385} and experimental~\cite{OC8434,PhysRevA.65.023806,16973851} work. In the limit of vanishing laser power, our measurements indicate that the EIT signal decreases exponentially as a function of detuning from the line center. This special behavior was theoretically predicted~\cite{PhysRevA.22.2115} for room-temperature atoms moving in Gaussian optical beams.

\section{Velocity-selective optical pumping and EIT}

\begin{figure}\label{fig2}
\includegraphics[width=\columnwidth]{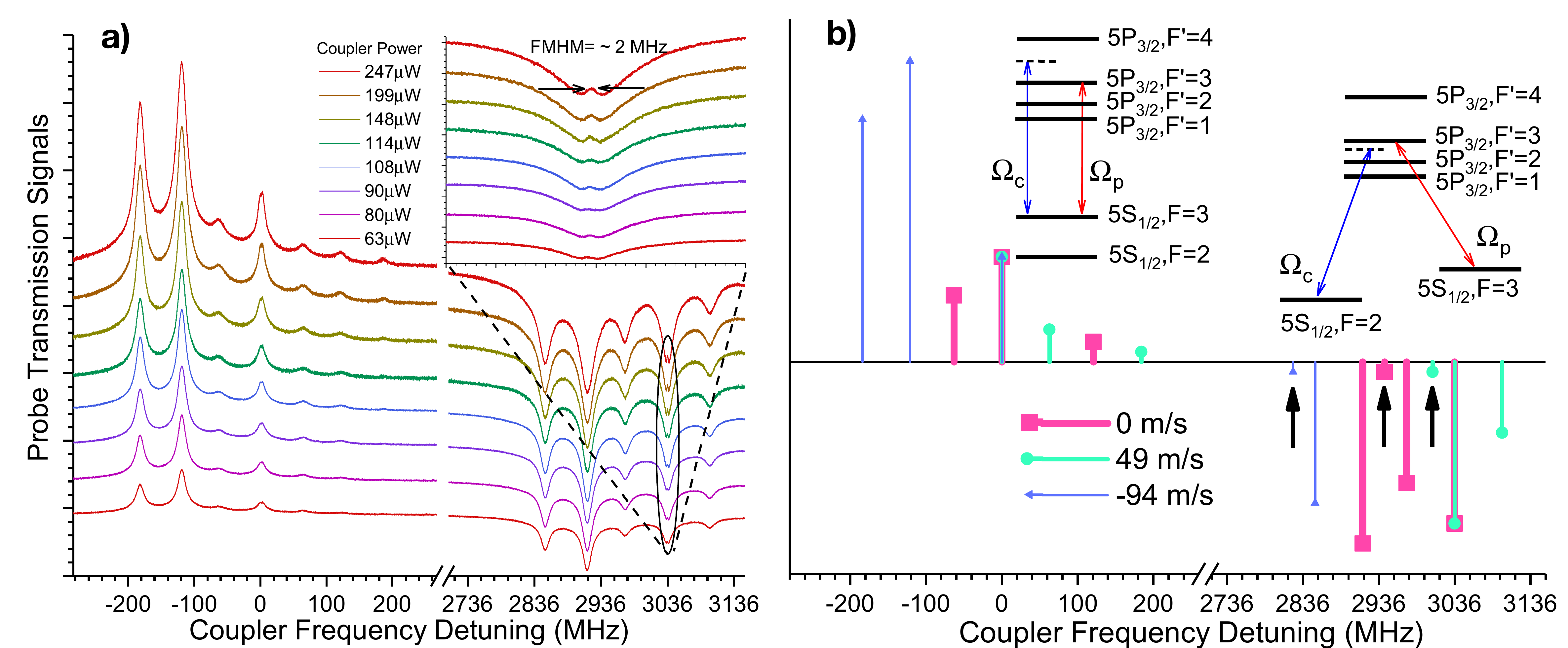}
\caption{(a) Series of probe transmission spectra with different coupler laser power vs coupler frequency detuning for fixed probe laser power (70~$\mu$W)  and fixed probe frequency tuned to the transition $\ket{5S_{1/2},F=3}\longrightarrow\ket{5P_{3/2},F'=3}$. The insert shows a zoom-in view on the EIT peak. (b) Analysis of the observed lines by atom velocity groups resonant with probe and coupler lasers.}
\end{figure}

\begin{table}
\caption{\label{table1}Assignment table for the spectral features observed in Figure~\ref{fig2}. 
The probe laser is fixed at the (zero-velocity) $\ket{5S_{1/2},F=3}\longrightarrow\ket{5P_{3/2},F'=3}$ transition frequency.
Left column: Resonant velocity groups for the indicated probe-laser transitions with lower- and upper-state hyperfine quantum numbers 
$[F_p, F']$. Center block:  Coupler-laser detunings of the enhanced-transmission peaks relative to the probe laser for the indicated coupler-laser transitions, with lower- and upper-state hyperfine quantum numbers $F_c=3$ and $F'$ and for the velocities shown in the left column. Right  block:  Coupler-laser detunings of the reduced-transmission peaks
relative to the probe laser for the indicated coupler-laser transitions, with lower- and upper-state hyperfine quantum numbers $F_c=2$ and $F'$ and for the velocities shown in the left column.}
\begin{tabular}{@{}*{7}{l}}
\br
\centre{1}{Probe Transition}&\centre{6}{Coupler Transition Detuning (MHz)}\\
\ns\ns\ns\ns
&\crule{6}\\
\ns\ns
Velocity Group &\centre{3}{From $\ket{F_c=3}$ to}&\centre{3}{From $\ket{F_c=2}$ to}\\
\ns
(m/s) ~ $[F_p, F']$ &$\ket{F'=2}$&$\ket{F'=3}$&$\ket{F'=4}$&$\ket{F'=1}$&$\ket{F'=2}$&$\ket{F'=3}$\\
\crule{1}&\crule{3}&\crule{3}\\
49 [3, 2] &0&63&184&3007&3036$^\mathrm{a}$&3009\\
0 [3, 3] &-63&0&121&2944&2973&3036$^\mathrm{a}$\\
-94 [3, 4] &-184&-121&0&2823&2852&2915\\
\br
\end{tabular}\\
\noindent  $^\mathrm{a}$ EIT Resonance\\
\end{table}

Velocity-selective effects occur in vapor cells because of the Doppler shift $\delta \omega = \vect{k}\cdot\vect{v}$, where $\vect{k}$ and $\vect{v}$ are optical wavenumber and atom velocity~\cite{doi:10.1080/09500340.2017.1328749}. It gives rise to velocity dependent ``hole burning'' (increased transmission peaks) and ``optical pumping'' (reduced transmission dips) effects demonstrated in the probe transmission spectra in Figure~\ref{fig2}(a). Both processes are highly velocity-selective due to the fact that the upper-state  ($\ket{5P_{3/2}}$) scattering rate scales as $s/[1+s+4(\Delta/\Gamma)^2]$, where $s$ is the saturation parameter defined as the ratio between laser and saturation intensity, $\Delta$ is the velocity-dependent optical detuning in rad/s, and $\Gamma$ is the natural decay rate, which is $2\pi \times6$~MHz for Rb 5$P_{3/2}$. Hence, at low saturation (our case) the velocity bandwidth of the D2 transition in a vapor cell is about 5~m/s. Figure~\ref{fig2}~(b) and Table~\ref{table1} relate the observed spectral lines to atomic transitions and resonant velocities. The line strengths vary due to the variation of transition dipole matrix elements between the hyperfine states, and because the resonances cover three velocity groups (with different values of the Maxwell probability distribution). Three resonances, indicated by the bold black arrows in Figure~\ref{fig2}(b), are too weak to become visible in Figure~\ref{fig2}(a). The line-strength ratios agree with a quantum Monte Carlo simulation~\cite{Zhang:18,Molmer:93,PhysRevLett.68.580}, in which we have included all magnetic sub-levels of the system. The ratios are not a main topic in the present paper and may be discussed in future work.

The insert of Figure~\ref{fig2}(a) shows the emergence of an EIT resonance on the optical-pumping dip centered at the hyperfine splitting 3036~MHz. The EIT results from quantum interference on two Raman-degenerate $\Lambda$ systems involving the excitation pathways $\ket{5S_{1/2},F=3}\leftrightarrow\ket{5P_{3/2},F'}$, driven by the probe laser, and $\ket{5S_{1/2},F=2} \leftrightarrow\ket{5P_{3/2}, F'}$, driven by the coupler, where $F'=2$ or 3. These couplings are velocity-selective in the Doppler-broadened medium; here, the respective resonant velocities are 0 and 49~m/s. The velocity difference is the smallest among $\Lambda$-EIT cases on the $^{85}$Rb and $^{87}$Rb D1 and D2 lines, and it is smaller than the thermal atom velocity in the cell. We find in Section~4 that EIT on the $^{85}$Rb D2 line is affected by both $\Lambda$-EIT systems.

\section{Zeeman shifts of EIT lines excited by phase locked lasers}

The line width of the EIT peak in Figure~\ref{fig2}(a) is about 2~MHz. Power broadening, relative laser frequency jitters and Zeeman shifts of the involved magnetic sub-levels due to stray magnetic fields are the dominant contributors to the line width. In the following experiments we have mitigated the last two broadening mechanisms by implementation of an Optical Phase Lock Loop (OPLL), and by application of a calibrated, longitudinal magnetic field that lifts the Zeeman degeneracies, allowing us to selectively study a magnetic-field-insensitive EIT resonance.

Atomic decoherence caused by laser frequency jitter~\cite{DALTON1982411,Dalton_1982} is significantly improved by OPLL~\cite{Benson10436771}. The measured power spectral density of the beat-note signal between our phased-locked lasers~\cite{Zhu:93} indicates a residual phase uncertainty $\sigma_{\phi_{rms}} < 0.3$~rad and a FWHM frequency width of less than 3~Hz. This result is comparable to Ref.~\cite{Benson10436771}, where  similar locking electronics is used.

\begin{figure*} \label{fig3}
\includegraphics[width=\columnwidth]{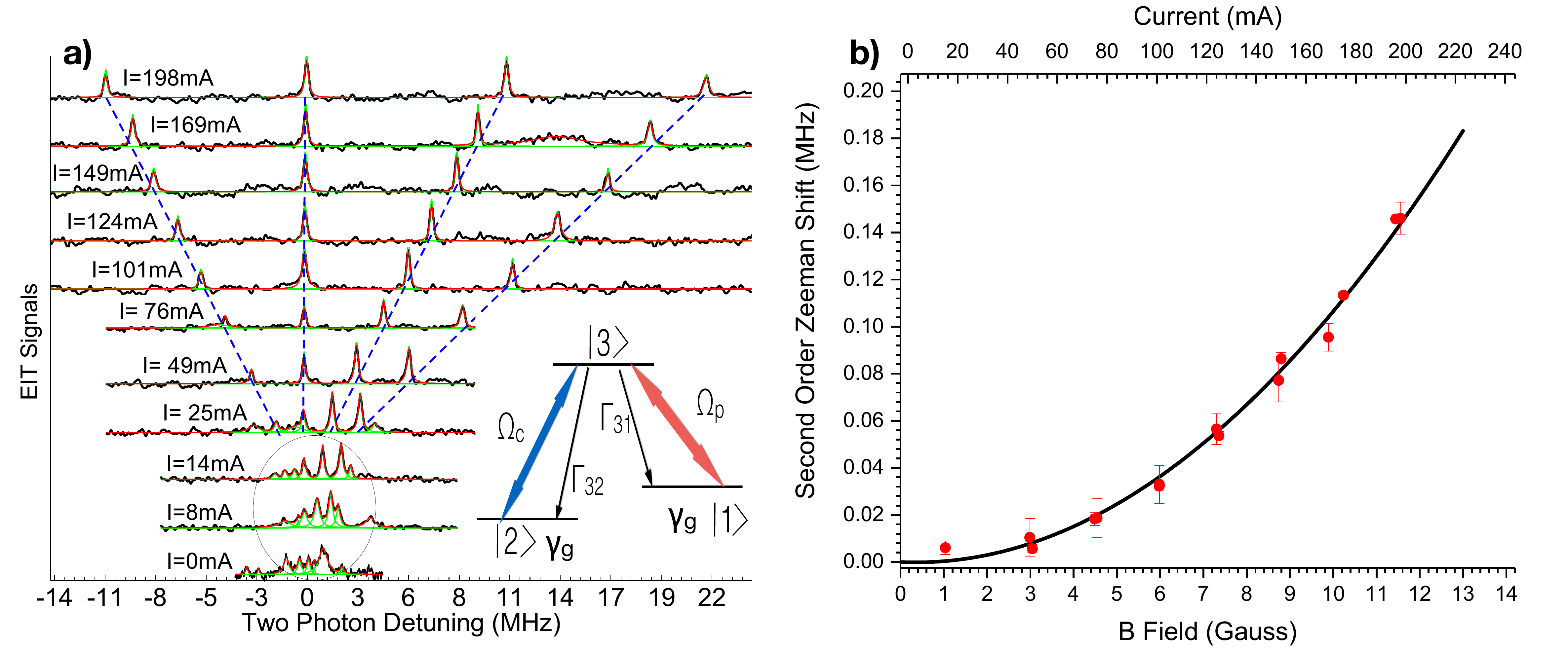}
\caption{
(a) Series of Zeeman-split EIT spectra (black lines) for different magnetic-coil currents. Fittings of individual peaks (green lines) and entire traces (red lines) are shown to guide the eye. The spectra are dominated by four Zeeman-split EIT lines, each of which corresponds with an isolated $\Lambda$ system (see insert). First-order Zeeman shifts (blue dashed lines) can be utilized to perform an in-situ calibration of the magnetic field vs current. In currents less than $\sim 25$~mA the spectra are affected by stray magnetic fields (circled region); this region is excluded from the field-calibration fit. (b) Measured (red dots) and theoretical (black line) second-order Zeeman shifts of the EIT resonance involving the $\ket{5S_{1/2}, F=2, m_F=-1}$ and $\ket{5S_{1/2}, F=3, m_F=1}$ ground states.
}
\end{figure*}

The EIT line broadening caused by stray magnetic fields~\cite{Huss:06} is alleviated by applying a comparably large, homogeneous magnetic field which removes degeneracies between the magnetic sub-levels. The Zeeman level diagram is shown in Figure~\ref{fig1}(b). In Figure~\ref{fig3}(a) we present the Zeeman shifts of the EIT signals. For the laser polarizations in our experiment, the first-order Zeeman shifts are $\delta_{Zeeman}=\mu_{B}B\left(-\frac{2}{3}+\frac{2}{3}m_F\right)$, where $\mu_B$ is the Bohr magneton, $B$ is the magnetic field, and $m_F$ is the magnetic quantum number of the state $\ket{5S_{1/2}, F=3, m_F}$. The three magnetic-field-sensitive EIT resonances ($m_F=\{0,2,3\}$) allow an in-situ calibration of $B$, against the coil current, $I$. The calibration factors for the field and the EIT line splittings are $61.7 \pm 0.8$Gauss/A and $57.6\pm0.7$~MHz/A, respectively. The uncertainty is obtained through a linear fitting procedure, which results in an $R^2$ value of 0.99993 with a confidence level of 99.5\%. In currents (fields) below $\sim 10$~mA ($0.6~$Gauss), the effects of transverse stray magnetic fields (circled region in Figure~\ref{fig3}(a)) become obvious.

The first order Zeeman shift vanishes for the EIT resonance involving the states $\ket{5S_{1/2}, F=2, m_F=-1}$ and $\ket{5S_{1/2}, F=3, m_F=1}$.  The minuscule shift of this EIT line due the second order Zeeman effect is plotted as red dots in Figure~\ref{fig3}(b). The black line is the expected second order Zeeman shift obtained through a direct diagonalization of the Hamiltonian including all magnetic sub-levels in both ground and excited states. At fields $B \gtrsim 3$~Gauss the EIT resonances become well-separated, and the magnetic-field-insensitive resonance becomes insensitive to line pulling and broadening effects. For the remainder of the paper, we choose a longitudinal field of $B=6$~Gauss. At this field strength, field variations due to the finite length of the solenoid and transverse stray fields are less than $1 \%$. The resultant variation of the second-order Zeeman shift causes inhomogeneous line broadening of $\lesssim 1.5~$kHz for the magnetic-field-insensitive EIT line.

\section{Doppler narrowing and beam Shape effects on EIT linewidth}

Taking advantage of the experimental techniques mentioned above, we are able to gain further insight into the ground-state decoherence in a $\Lambda$ system by studying the line width of the magnetic-field-insensitive EIT resonance. In the limit of zero Rabi frequency, the line width is limited by collision~\cite{PhysRevA.63.043813,PhysRevA.58.2345,Knappe:01} and transit-time effects~\cite{PhysRevA.22.2115}, in addition to technical noise such as residual relative phase fluctuations of the lasers~\cite{PhysRevA.63.043813} and stray magnetic fields caused by the coil current noises. For experiments using buffer-gas-free room-temperature vapor cells, collisions between Rb atoms and other trace gas atoms are less important. Therefore, power broadening, transverse laser intensity distribution, and transit-time effects of the thermal atoms become the major factors, as we demonstrate in the following. In addition, wall collisions are still present which deplete ground-state coherence~\cite{15108061}.

\begin{figure} \label{fig4}
\includegraphics[width=\columnwidth]{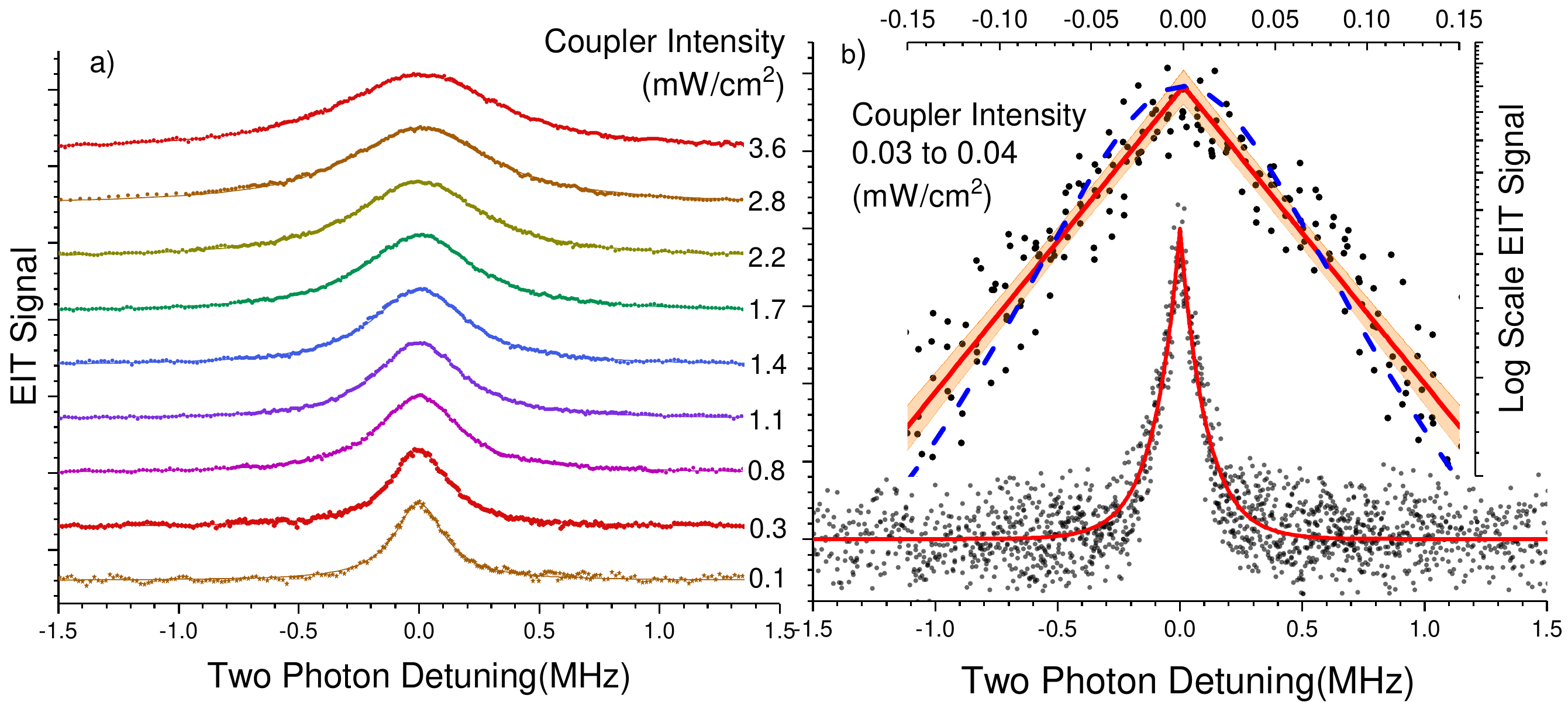}
\caption{ 
(a) EIT resonances for the indicated coupler-laser intensities at the beam center. Experimental data, shown as dots, are fit very well by Lorentzians (solid curves). (b) EIT line shape in the limit of very small coupler laser intensity. Several data sets for intensities ranging from 0.03 to 0.04~mW/cm$^2$ are overlapped (black dots) in order to improve statistics. The red solid curve represents a symmetric exponential-decay fit on both sides of the resonance. A log-scale representation of data and fits are shown as an insert, with the $99.5\%$-confidence range shown in orange. A Lorentzian fit (blue dashed curve), plotted for comparison, clearly is less good.
}
\end{figure}

Figure~\ref{fig4}(a) shows a series of spectra of the magnetic-field-insensitive  EIT resonance vs coupler-laser intensity at $B= 6$~Gauss. At higher intensities power broadening dominates, and the EIT lines have a Lorentzian shape (as opposed to Gaussian or symmetric exponential). As the intensity drops below $\sim 0.1~$mW/cm$^2$, the line width drops dramatically, and the line shape deviates from a Lorentzian profile. Figure~\ref{fig4}(b) shows spectra with intensities between 0.03 and 0.04~mW/cm$^2$. These low-intensity signals show an exponential decay on both sides of the resonance. This special behavior has been predicted theoretically in Ref.~\cite{PhysRevA.22.2115} as a consequence of thermal atoms traveling through Gaussian optical beams. Due to limited signal to noise ratio, we are not able to resolve the exact second derivative at the line center. It needs to be pointed out that this transit-time effect is fundamentally different from CPT line shapes observed in buffer-gas-enriched vapor cells. In the latter case, the diffusion~\cite{PhysRevLett.96.043601,8698263} of the alkali atoms among the buffer gas atoms and the local intensity~\cite{PhysRevA.69.024501} of the driving laser beams play dominant roles.

\begin{figure}\label{fig5}
\includegraphics[width=0.7\columnwidth]{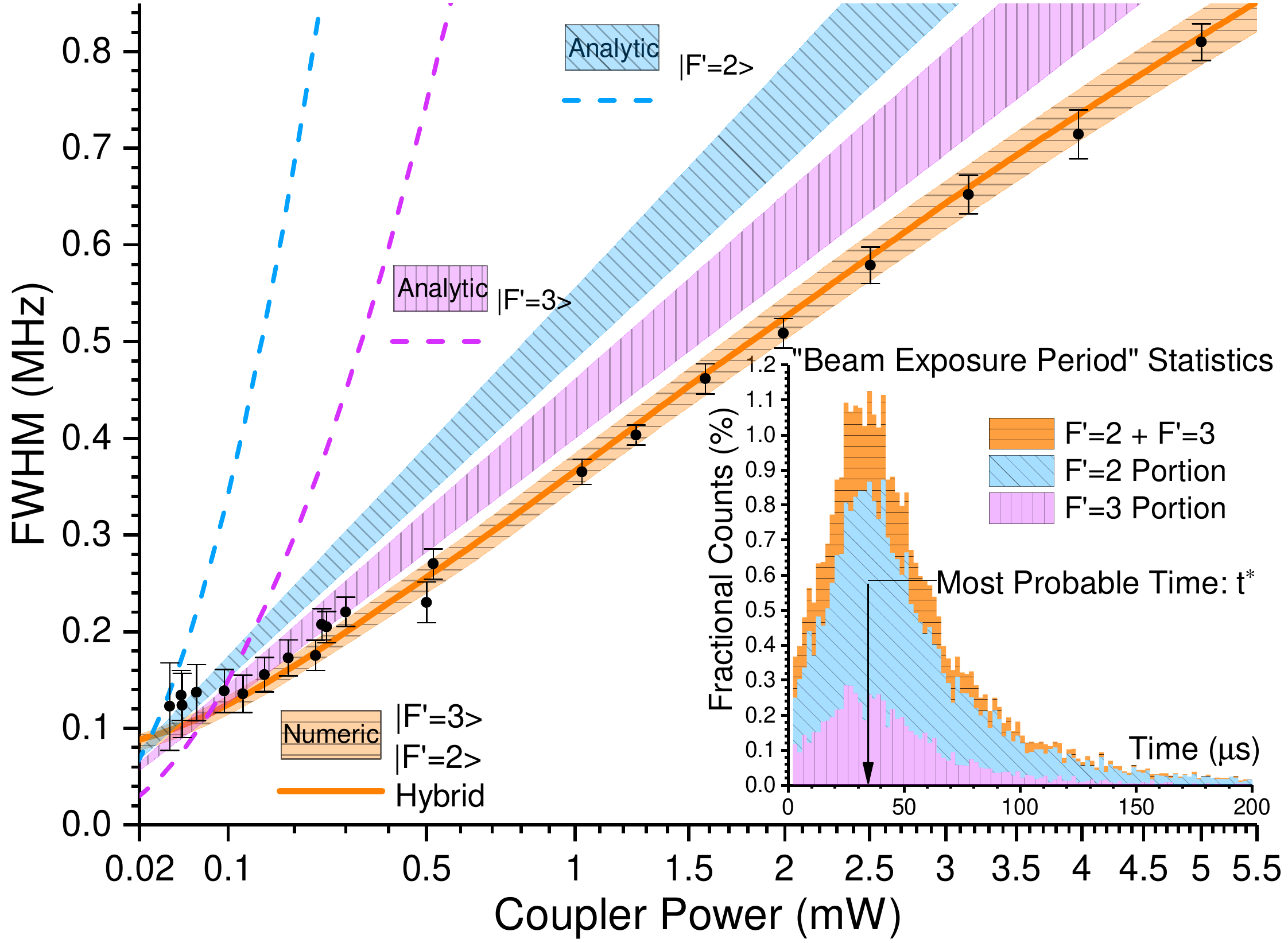}
\caption{
Full widths at half maximum (FWHM) of EIT lines (black dots) as a function of coupler laser power. The length along the x-axis is linear in square-root of power. Blue and purple dashed lines represent results for stationary atoms.  The blue and purple hatched areas are analytic results following Ref.~\cite{PhysRevA.66.013805,7653385} for 300-K thermal atoms, calculated for single three-level $\Lambda$ systems involving the excited states $F'=2$ (blue, diagonally hatched) or 3 (purple, vertically hatched). The ground state decoherence rate $\gamma_{g}/2\pi$ is varied from 30~kHz to 40~kHz over the shaded hatched regions. The different trends are due to different Rabi frequencies from the beam center for a given laser power and beam size (see text for detail). The orange curve is a simulation result (see Appendix for details) in which both Raman-degenerate $\Lambda$ systems with $F'=2$ and 3 are accounted for, as described in the text. The variation of the Rabi frequencies transverse to the beam directions is included, and $\gamma_{g}/2\pi$ is 35~kHz. The orange-shaded, horizontally hatched region represents a sweep of $\gamma_{g}/2\pi$, the only fitting parameter in the model, from 30~kHz to 40~kHz. The insert shows a histogram of the ``Beam Exposure Period (BEP)'' defined in text; the bin size is $2~\mu$s and the most probable BEP is $t^* = 33~\mu$s.
}
\end{figure}

In Figure ~\ref{fig5}, we plot the measured FWHM (black dots) as a function of the coupler laser power and compare to various analytic and numeric models. We note first that the measured line width is much lower than an opacity/density adjusted result~\cite{PhysRevLett.79.2959,RevModPhys.77.633} (blue and purple dashed lines) for a homogeneously broadened sample, such as cold atoms or thermal vapor cells with buffer gases~\cite{PhysRevA.58.2345,PhysRevA.63.043813}, which clearly does not apply to our case. An analytic result for a single $\Lambda$ system in a Doppler-broadened system,  given in Ref.~\cite{PhysRevA.66.013805,7653385}, reproduces the general trends in our data (blue and purple hatched areas) in terms of approximate line width values as well as the linear scaling of the width with Rabi frequency (which is linear in distance along the x-axis in Figure~\ref{fig5}). The remaining mismatch between the analytic result and our measurements, together with the exponential-decay-like line shape (Figure~\ref{fig4}(b)), motivate us to investigate the effects caused by (1) the presence of two Raman-degenerate $\Lambda$ systems and (2) by the transverse inhomogeneity of the laser intensities (thus the Rabi frequencies) away from the beam axes.

Before discussing the EIT line width in more detail, we recall that in the $B=0$ analysis two Raman-degenerate $\Lambda$ configurations involving two different velocity classes contribute to the EIT signal. This remains true at $B=6$~Gauss, with the velocity groups resonantly coupled to states $\ket{5P_{3/2}, F', m_F=0}$ differing by about 50~m/s for $F'=2$ and 3. Since this velocity difference is much less than the RMS thermal velocity of 170~m/s in one dimension, both $\Lambda$ configurations contribute to the EIT line and its width. Further, angular matrix elements and Rabi frequencies depend strongly on magnetic field due to the onset of hyperfine de-coupling in the excited state. At 6~Gauss and for the given circular polarizations, the angular matrix elements, $w_{i,F'}$, are $0.2630$ and $0.5825$ for probe and coupler transitions resonant with $\ket{5P_{3/2},F'=2,m_F=0}$, respectively. For probe and coupler resonant with  $\ket{5P_{3/2},F'=3,m_F=0}$, they are $0.4140$ and $0.3323$. The Rabi frequencies are then given by $\Omega_{i,F'} = \Gamma \sqrt{I_i/(2 I_{sat})} \times w_{i,F'}$,with $\Gamma = 2 \pi \times 6$~MHz and $I_{sat}=1.67$~mW/cm$^2$. Subscript $i$ stands for probe or coupler. These Rabi-frequency expressions are used in Figure~\ref{fig5}, with the given beam powers and widths.

In our numerical model, we integrate three-level Lindblad equations for an ensemble of atom trajectories with random initial velocities, drawn from a 3D Maxwell distribution, and random initial positions on the cell walls or windows. Since the two Raman-degenerate $\Lambda$ EIT resonances, mentioned above, are only a few m/s wide in velocity space, for any given atom trajectory we select the upper-state $F'$-level that is closer to resonance with the probe laser, in the frame of reference of the moving atom, and solve the three-level Lindblad equation with that $F'$ level. In addition, the transverse laser intensity distributions are accounted for via spatially dependent Rabi frequencies. Also, the vapor opacity in our experiment is kept at a sufficiently low value that the longitudinal intensity variation of the beams, caused by absorption, can be neglected. A more detailed description can be found in the Appendix. In this simulation, the ground-state decoherence rate, $\gamma_{g}$, is the only fitting parameter.  As shown in Figure~\ref{fig5}, the numerical simulation (solid orange line) fits our data very well for $\gamma_{g}/2\pi=35$~kHz, with a confidence range of about 30~kHz to 40~kHz (orange hatched area).

\section{Discussions and Conclusions}

We note that in Ref~\cite{PhysRevA.66.013805,7653385} the decay of the coherence $\rho_{12}$ is modeled via a bidirectional symmetric population transfer rate between the two ground states, $\vert 1 \rangle$ and $\vert 2 \rangle$. This mechanism is useful to describe open systems where atoms move in and out of an interaction region~\cite{7653385,7633247}, and it may also be used to describe the population decay due to atom-wall collisions. In our work we adopt the model from Ref~\cite{RevModPhys.77.633}, where the decay of the coherence $\rho_{21}$ is modeled through dephasing only, while the ground-state population exchange occurs exclusively via optical pumping through the excited state (for details see Appendix).

In our next discussion point, we draw a distinction between dephasing of $\rho_{21}$ and interaction-time broadening. Both effects are ubiquitous in thermal-gas experiments. An analytic approach can be found in Ref.~\cite{PhysRevA.22.2115}. Comparing to experiments using atomic beams or cold atoms, interaction time in the vapor cell can be thought of as the beam exposure period ("BEP") i.e. the time of flight of the atoms through the laser-beam core, defined as the region with diameter $2w_p$ and length of $L$, where $w_p=6.5$~mm is the usual 1/e-dropoff radius of the electric field in our Gaussian beams, and $L=70$~mm is the length of the vapor cell. The BEP is broadly distributed due to beam and cell geometry, randomness in atom velocity, and randomness in trajectory orientation relative to cell and beams. Here, the most probable value of the BEP $t^*=33~\mu$s (see insert of Figure~\ref{fig5}). Over a range of numerical tests we have seen that $t^* \approx a \times(2w_p/u)$, with a numerical constant $a=0.51$ and the most probable speed for a 3D Maxwell velocity distribution $u=\sqrt{2k_{B}T/m_{Rb}}$. The tests have also shown that $a$ depends somewhat on the geometric ratios $w_p/R$ and $R/L$; it varies by about 20\% from the quoted value for $w_p/R$ varying from 0.2 to 0.8 and $R/L$ from 0.1 to 0.5.

According to our numerical model, the zero-power line width is about $2\gamma_{g} \approx 2\pi\times70$~kHz. Interaction-time broadening, which is on the order of $1/t^*\approx30$~kHz, is included in our simulation in Figure~\ref{fig5} and has a relatively minor effect on the simulated zero-power line width. The lowest line width experimentally measured, about 100~kHz, is still slightly affected by power broadening. It is noted that the experimental uncertainty bars in Figure~\ref{fig5} increase at low powers due to the decrease in photo-current. Even at the lowest powers, experimental and simulated line widths agree within the experimental uncertainty.

The question arises where the decoherence $\gamma_g$ comes from. Decoherence due to the spin exchange collisions between Rb atoms is an unlikely cause, as it is only on the order of tens of Hz~\cite{RevModPhys.44.169} at our vapor density (about $10^{10}\mathrm{cm}^{-3}$). Also, differential phase noise between coupler and probe lasers is an unlikely cause, because the residual phase noise of the OPLL is only 0.3~rad, and the spectral width of the laser beat signal at 3~GHz has been directly measured to be below about 3~Hz.

Looking at other causes, we note that recent spin noise measurements of Faraday rotation signals~\cite{Sekiguchi,PhysRevA.101.013821} carried out in buffer-gas free Rb vapor cells have revealed that the ground state $1/T_2$ rate can vary from kHz to hundreds of kHz, depending on whether the cell walls are coated with anti-relaxation layers or not. Models provided in Ref.~\cite{PhysRevA.101.013821} also suggest that as low as a few mTorr background gas, which can either come from the outgassing of the coating layer or an impurity introduced during cell manufacturing, can reduce the mean free path of the Rb atoms from meters (much larger than practical cell size) to millimeters (which is on the order of typical optical beam sizes). Since the effects of collisional interactions on quasi-steady-state EIT spectra are not covered in our ballistic model, while wall interactions are effectively included via the BEP time limitation and a random initialization of the ground state population distribution before the atoms desorb from the wall/window, we speculate that the decoherence measured in our work may originate in collisions with an impurity gas.

In conclusion, we have explored $\Lambda$ EIT in a Rb vapor cell on the D2 line as a means to study EIT line-width suppression in a Doppler-broadened medium. Lifting Zeeman degeneracies by application of a homogeneous magnetic field of $6~$Gauss has allowed us to focus the study on a single, magnetic-field-insensitive EIT line, and to push our study of EIT line width vs beam intensity into the 100-kHz regime. We have qualitatively explained the EIT line width behavior using existing analytical models and achieved quantitative agreement using a numerical approach in which we have included experimentally relevant details. We have observed a remaining ground-level dephasing rate $\gamma_g/2\pi \sim 35$~kHz that could not be readily explained. We have discussed possible causes for $\gamma_g$. In this context, one may explore the $\Lambda$ EIT line width as a measure to analyze residual gases in closed cells, where tools such as residual gas analyzers cannot be used. In future, improved models may be developed to account for effects introduced by optical pumping and atomic decay~\cite{Zhang:18, PhysRevA.99.053426} among all magnetic sublevels in both $\ket{5S_{1/2}}$ and $\ket{5P_{3/2}}$ hyperfine manifolds. Effects induced by state mixing via transverse magnetic fields and impurities in laser polarization states and frequency spectra may also be included.

\ack
This work was supported by NSF Grant No.~1707377. We thank Dr. David A. Anderson of Rydberg Technologies, Inc. and Prof. Eric Paradis for valuable discussions. We also thank Prof. Alex Kuzmich and Prof. Duncan Steel for lending us lab equipment.

\appendix
\section*{Appendix: Numeric Modeling}
\setcounter{section}{1}

A three level $\Lambda$-type model is implemented with  $\ket{5S_{1/2},F=3,m_F=1}$ as state $\ket{1}$, $\ket{5S_{1/2},F=2,m_F=-1}$ as state $\ket{2}$, and $\ket{5P_{3/2}, F', m_F=0}$ as state $\ket{3}$. Atoms move on trajectories $\vect{r}(t) = \vect{r}_0 +\vect{v}_0t$ with initial random  velocities $\vect{v}_0$ from a 3D Maxwell distribution, and initial positions $\vect{r}_0$ randomly chosen on cell walls/windows. States $\ket{1}$ and $\ket{3}$ are coupled by a position-dependent probe Rabi frequency $\Omega_p(\vect{r}(t))$, and states $\ket{2}$ and $\ket{3}$ by a coupler  Rabi frequency $\Omega_c(\vect{r}(t))$. The system has two sets of $\Lambda$ couplings, one for $F'=2$ and another for  $F'=3$. For each atom of the ensemble, the $F'$-value in state $\ket{3}$ is picked such that the Doppler shift of the EIT lasers in the atom's rest frame is minimized for the atom's ${\bf{v}}_0$-value. This is allowed because the internal-state dynamics is usually dominated by the $\Lambda$ system the atom is closer in resonance with.  

The position-dependent Rabi frequencies $\Omega_{c,p}$ are given by $\Omega_{c,p}(\vect{r})=\vect{\mu}_{ij}\cdot\vect{E}_{c,p}(\vect{r})/\hbar$ where $\vect{\mu}_{ij}$ is the transition electric dipole moment between state $\ket{i}$ and $\ket{j}$, and $\vect{E}_{c,p}(\vect{r})$ are electric fields with Gaussian transverse profiles. The dipole moments $\vect{\mu}_{ij}$ are obtained by diagonlization of the atomic Hamiltonian with all Zeeman and hyperfine interactions included. The dipole moments depend significantly on the magnetic field. At $B=6~$Gauss and for the laser polarizations used, for $F'=2$ it is $\mu_{31}=1.46ea_0$ and  $\mu_{32}=3.23ea_0$, and for $F'=3$ it is $\mu_{31}=2.30ea_0$, $\mu_{32}=1.84ea_0$.

In the two-color field picture (which is applicable to systems with fields of sufficiently different frequencies), the atom-laser interaction Hamiltonian in the space $\{ \ket{1}, \ket{2}, \ket{3} \}$  is

\begin{equation}
H_{int}=-\frac{\hbar}{2}
        \left[
		\matrix{
		0 & 0 & \Omega_p(\vect{r}) \cr
		0 & -2(\Delta_1-\Delta_2) & \Omega_c(\vect{r}) \cr
		\Omega_p(\vect{r}) & \Omega_c(\vect{r}) & -2\Delta_1
		}
		\right]
\end{equation}
where $\Delta_1=\omega_p-\omega_{31}-\vect{k}_p \cdot \vect{v}_0$ and $\Delta_2=\omega_c-\omega_{32}-\vect{k}_c \cdot \vect{v}_0$ are the velocity-dependent detunings of the fields relative to the atomic transition frequencies $\omega_{ij}$.

The dynamics of the laser-driven atomic system is described by the Lindblad equation for the density operator $\rho$,

\begin{eqnarray}
\frac{\mathrm{d}\rho}{\mathrm{d}t} =
    & \frac{1}{i\hbar}[H_{int},\rho] \cr
	&+\frac{\Gamma_{31}}{2}[2\hat{\sigma}_{13}\rho\hat{\sigma}_{31},-\hat{\sigma}_{33}\rho-\rho\hat{\sigma}_{33}]\cr
	&+\frac{\Gamma_{32}}{2}[2\hat{\sigma}_{23}\rho\hat{\sigma}_{32},-\hat{\sigma}_{33}\rho-\rho\hat{\sigma}_{33}]\cr
	&+\frac{\gamma_{1}}{2}[2\hat{\sigma}_{11}\rho\hat{\sigma}_{11},-\hat{\sigma}_{11}\rho-\rho\hat{\sigma}_{11}]\cr
	&+\frac{\gamma_{2}}{2}[2\hat{\sigma}_{22}\rho\hat{\sigma}_{22},-\hat{\sigma}_{22}\rho-\rho\hat{\sigma}_{22}]\cr
	&+\frac{\gamma_{3}}{2}[2\hat{\sigma}_{33}\rho\hat{\sigma}_{33},-\hat{\sigma}_{33}\rho-\rho\hat{\sigma}_{33}]
\end{eqnarray}

with atomic projection operators $\hat{\sigma}_{ij}=\ket{i}\bra{j}$,  dephasing rates $\gamma_1$, $\gamma_2$ and $\gamma_3$, and partial spontaneous decay rates $\Gamma_{31}$ and $\Gamma_{32}$. The latter, within the Weisskopf-Wigner approximation, are given by~\cite{berman_malinovsky_2011},

\begin{equation}
\fl \Gamma_{3i} =\frac{4}{3}\alpha_{\mathrm{FS}}\frac{1}{2F_3+1}\frac{\omega_{3i}^3}{c^2}(2F_3+1)(2F_i+1)\left[\langle 5P_{3/2}\|r\|5S_{1/2}\rangle
                        \left\{
						    \matrix
						    {
							J_3 & 1 & J_i \cr
							F_i & I & F_3
					    	}
						\right\}
						\right]^2
\end{equation}

where $\alpha_{FS}$ is the fine structure constant, $F_i$ and $J_i$ are the  F and J quantum numbers of state $\ket{i}$,  $\langle 5P_{3/2}\|r\|5S_{1/2}\rangle=\sqrt{2J_i+1}\times4.23a_0$ is the reduced dipole matrix element of the D2 transition of $^{85}$Rb, and $\{*\}$ represents the Wigner-6J symbol. Using this equation, $\Gamma_{31}=2\pi\times1.35$~MHz and $\Gamma_{32}=2\pi\times4.72$~MHz for state $\ket{3}=\ket{5P_{3/2}, F'=2, m_F=0}$, and $\Gamma_{31}=2\pi\times 3.37$~MHz and $\Gamma_{32}=2\pi\times 2.70 $~MHz for  $\ket{3}=\ket{5P_{3/2}, F'=3, m_F=0}$. The total spontaneous decay rate $\Gamma_3$ of state $\ket{3}$ is $\Gamma_3=\Gamma_{31}+\Gamma_{32}=2\pi\times6.07$~MHz, the natural decay rate of Rb 5$P_{3/2}$.

The decoherence rate $\gamma_3$ is dominated by laser-frequency noise. The lasers are locked via standard saturation-absorption-spectroscopy, with an estimated $\gamma_3 \sim 2 \pi \times 200$~kHz. The exact value is not important because $\gamma_3$ does not affect the line width of the EIT signal~\cite{RevModPhys.77.633}. 

For simplicity, we set $\gamma_1=\gamma_2=\gamma_g$ in our discussion. The decoherence rate $\gamma_g$ includes noise on the frequency difference of coupler and probe lasers and collisional ground-state level dephasing. The former is very small, due to our use of an OPLL, while the latter could be several tens of kHz due to collisions between Rb atoms and cells walls or trace gases inside the cell. Here we find a fitted $\gamma_g \approx 2 \pi \times 35$~kHz. 

We numerically integrate the Lindblad equation for a large ensemble of trajectories with randomly chosen velocities $\vect{v}_0$ and initial positions $\vect{r}_0$, as explained above. The initial populations are set to be randomly distributed between states $\ket{1}$ and $\ket{2}$, with $\rho_{11}(t=0)+\rho_{22}(t=0)=1$. The position-dependence of the Rabi frequencies, $\Omega_{c,p}(\vect{r})$, enters in the time integration via the atom trajectories, $\vect{r}(t) = \vect{r}_0 +\vect{v}_0 t$. The integration for a given atom ends when its trajectory exits the cell volume (i.e., hits a wall/window). The absorption signal and the EIT then follow 

\begin{equation}
 \frac{1}{N}\sum_{j=1}^{N}\int_{0}^{T_j}{\rm{Im}} [\rho_{31}(t;j)]\mathrm{d}t
\end{equation}
where $j$ is a trajectory label, $N$ the number of trajectories, and $T_j$ the time of flight of atom $j$ through the cell. 

\section*{References}

\bibliographystyle{iopart-num}
\bibliography{lambdaEIT}

\end{document}